\documentclass{aa}
\usepackage[varg]{txfonts}
\usepackage{graphicx}
\usepackage{natbib,twoopt}
\usepackage{hyperref} %% to avoid \citeads line fills
\bibpunct{(}{)}{;}{a}{}{,} %% natbib format for A&A and ApJ
\begin{document}

   \title{Waiting to make an impact: A probable excess of near-Earth asteroids in 
          2018~LA-like orbits
         }
   \author{C. de la Fuente Marcos$^{1}$
           \and
           R. de la Fuente Marcos$^{2}$}
   \authorrunning{C. de la Fuente Marcos \and R. de la Fuente Marcos}
   \titlerunning{An excess of 2018~LA-like orbits 
                 }
   \offprints{C. de la Fuente Marcos, \email{nbplanet@ucm.es}
              }
   \institute{$^1$ Universidad Complutense de Madrid,
              Ciudad Universitaria, E-28040 Madrid, Spain \\
              $^2$AEGORA Research Group,
              Facultad de Ciencias Matem\'aticas,
              Universidad Complutense de Madrid,
              Ciudad Universitaria, E-28040 Madrid, Spain
              }
   \date{Received 24 September 2018 / Accepted 28 November 2018}

   \abstract
      {The discovery and tracking of 2018~LA marks only the third instance in 
       history that the parent body of a fireball has been identified before 
       its eventual disintegration in our atmosphere. The subsequent recovery 
       of meteorites from 2018~LA was only the second time materials from 
       outer space that reached the ground could be linked with certitude to a 
       particular minor body. However, meteoroids like 2018~LA and its 
       forerunners, 2008~TC$_{3}$ and 2014~AA, are perhaps fragments of larger 
       members of the near-Earth object (NEO) population. As the processes 
       leading to the production of such fragments are unlikely to spawn just 
       one meteoroid per event, it is important to identify putative siblings 
       and plausible candidates from which the observed meteoroids might have 
       originated. 
       }
      {Here, we study the pre-impact orbital evolution of 2018~LA to place 
       this meteoroid within the dynamical context of other NEOs that follow 
       similar trajectories. 
       }
      {Our statistical analyses are based on the results of direct $N$-body 
       calculations that use the latest orbit determinations and include 
       perturbations by the eight major planets, the Moon, the barycentre of 
       the Pluto--Charon system, and the three largest asteroids. A 
       state-of-the-art NEO orbit model was used to interpret our findings and 
       a randomization test was applied to estimate their statistical 
       significance.
       }
      {We find a statistically significant excess of NEOs in 2018~LA-like 
       orbits; among these objects, we find one impactor, 2018~LA, and the 
       fourth closest known passer-by, 2018~UA. A possible connection with the 
       $\chi$-Scorpiids meteor shower is also discussed. The largest known NEO 
       with an orbit similar to that of 2018~LA is the potentially hazardous 
       asteroid (454100) 2013~BO$_{73}$ and we speculate that they both 
       originate from a common precursor via a collisional cascade. 
       }
      {Future spectroscopic observations of 454100 and other NEOs in similar 
       orbits may confirm or deny a possible physical relationship with 
       2018~LA. 
       }

         \keywords{methods: numerical --
                   meteorites, meteors, meteoroids -- 
                   minor planets, asteroids: general --
                   minor planets, asteroids: individual: (454100) 2013~BO$_{73}$ --
                   minor planets, asteroids: individual: 2018~LA  --
                   minor planets, asteroids: individual: 2018~UA
                  }

   \maketitle

   \section{Introduction}
      Planet-crossing asteroids may deliver meteorites to the inner planets when their paths intersect with those of the planets at the 
      proper times, leading to an impact (see e.g. \citealt{1998M&PS...33..999M}). On the Earth, meteorites collected after the 
      observation of bright fireballs or bolides have sometimes been tentatively traced back to the near-Earth object (NEO) 
      population or collisional events in the main asteroid belt (see e.g. \citealt{2007MNRAS.382.1933T}, \citeyear{2015MNRAS.449.2119T}; 
      \citealt{2015MNRAS.454.2965O}; \citealt{2018Icar..311..271G}). However, only in two cases, those of 2008~TC$_{3}$ 
      (\citealt{2009Natur.458..485J}) and 2018~LA,\footnote{\href{https://seti.org/fragment-impacting-asteroid-recovered-botswana-0}
      {Fragments found in Botswana's Central Kalahari Game Reserve.}} the parent body of the meteor airburst that produced the 
      recovered meteorites was observed prior to entering the atmosphere of our planet. In fact, just three natural objects have 
      been discovered before colliding with the Earth: the two already mentioned and 2014~AA (\citealt{2016Ap&SS.361..358D}; 
      \citealt{2016Icar..274..327F}), but the impact of 2014~AA took place over the Atlantic Ocean, far from any land masses. All three were 
      small Apollo asteroids that might well have been fragments of larger objects. In the event of having been fragments, identifying their 
      parent bodies could be quite difficult as they tend to follow rather chaotic pre-impact orbits (see e.g. \citealt{2016Ap&SS.361..358D}) 
      and perhaps be the result of collisional cascades that produce multi-generation fragments, that is, fragments of fragments (see e.g. 
      \citealt{1994IAUS..160..205F}).

      In principle, meteoroids like 2008~TC$_{3}$, 2014~AA, or 2018~LA can be spawned by another small body from a larger object 
      during subcatastrophic collisions (see e.g. \citealt{2007Icar..186..498D}), but they can also be released as a result of the 
      Yarkovsky--O'Keefe--Radzievskii--Paddack (YORP) mechanism (see e.g. \citealt{2006AREPS..34..157B}) or out of tidal 
      disruption events during very close encounters with planets (see e.g. \citealt{2014Icar..238..156S}). Here, we investigate 
      the existence of a possible excess of NEOs in orbits similar to that of 2018~LA, parent body of the fireball observed over 
      South Africa and Botswana on 2018 June 2. This paper is organized as follows. In Sect. 2, we present the data used and 
      provide the details of our numerical model. Section 3 investigates the existence of putative siblings of 2018~LA and of 
      plausible candidates from which meteoroids like 2018~LA might have originated. Observational evidence on periodic meteor 
      activity perhaps linked to 2018~LA is presented in Sect. 4. In Sect. 5, we compare with predictions from a new 
      four-dimensional model of the NEO population. Our results are discussed in Sect. 6 and our conclusions summarized in 
      Sect. 7. 

   \section{Meteoroid 2018 LA: data and tools}
      Asteroid 2018~LA was first observed on 2018 June 2 by R.~A. Kowalski using the 1.5-m telescope of the Mount Lemmon Survey in 
      Arizona (\citealt{2018MPEC....L...04K}) and it was soon 
      determined\footnote{\href{https://archive.fo/20180602223233/https://www.projectpluto.com/neocp2/mpecs/ZLAF9B2.htm}
      {``Pseudo-MPEC" for ZLAF9B2 = 2018 LA by Bill Gray.}} that it was on a probable collision course 
      with our planet. As a boulder-sized object, it was expected to disintegrate in the atmosphere, well above the surface of the 
      Earth, producing a bright fireball. The fireball was indeed observed (\citealt{2018MPEC....L...04K}) and consistent 
      infrasound signals were detected by the sensors of the Comprehensive Nuclear-Test-Ban Treaty Organization (CTBTO, Mialle, 
      2018, priv. comm.).

      The orbit determination of 2018~LA available from Jet Propulsion Laboratory's (JPL) Small-Body Database
      (SBDB)\footnote{\href{https://ssd.jpl.nasa.gov/sbdb.cgi}{https://ssd.jpl.nasa.gov/sbdb.cgi.}} has been updated several times 
      since the discovery of the object and the latest solution (computed by D. Farnocchia on 2018 July 24) is based on 15 data 
      points ---14 optical observations and one fireball report. Table \ref{elements}, second column from the right, shows the 
      latest orbit determination released by JPL's SBDB. A preliminary exploration of the pre-impact orbital evolution of 2018~LA 
      using two previous orbit determinations (also published by JPL's SBDB) can be found in \citet{2018RNAAS...2b..57D} and 
      \citet{2018RNAAS...2c.131D}. In the following we only consider the latest one, which is probably and currently the most 
      accurate, publicly available orbit determination of 2018~LA.     
%
%------------------------------------------------------------------------------------------------------------------------- TABLE I
%---------------------------------------------------- Orbital elements asteroids (454100) 2013 BO73, 2016 LR, 2018 BA5 and 2018 LA 
%
      \begin{table*}
         \centering
         \fontsize{8}{11pt}\selectfont
         \tabcolsep 0.15truecm
         \caption{\label{elements}Heliocentric Keplerian orbital elements of asteroids (454100) 2013~BO$_{73}$, 2016~LR, 
                  2018~BA$_{5}$, 2018~LA, and 2018~TU. 
                 }
         \begin{tabular}{ccccccc}
            \hline\hline
             Parameter                                         &   &   454100                      &   2016~LR           & 2018~BA$_{5}$       & 
                                                                       2018~LA                     &   2018~TU            \\              
            \hline
             Semi-major axis, $a$ (AU)                         & = &   1.331499402$\pm$0.000000013 &   1.3812$\pm$0.0009 &   1.3671$\pm$0.0009 &  
                                                                       1.3763$\pm$0.0003           &   1.3782$\pm$0.0007  \\
             Eccentricity, $e$                                 & = &   0.41829734$\pm$0.00000004   &   0.4351$\pm$0.0005 &   0.4308$\pm$0.0005 &   
                                                                       0.4318$\pm$0.0002           &   0.4373$\pm$0.0004  \\
             Inclination, $i$ (\degr)                          & = &   4.543490$\pm$0.000006       &   2.542$\pm$0.002   &   4.538$\pm$0.007   & 
                                                                       4.2968$\pm$0.0013           &   4.071$\pm$0.004    \\
             Longitude of the ascending node, $\Omega$ (\degr) & = &  24.80181$\pm$0.00003         &  98.059$\pm$0.005   & 338.44$\pm$0.02     &  
                                                                      71.86963$\pm$0.00003         & 188.3899$\pm$0.0004   \\ 
             Argument of perihelion, $\omega$ (\degr)          & = & 298.10992$\pm$0.00004         & 236.508$\pm$0.007   &  42.65$\pm$0.02     & 
                                                                     256.0482$\pm$0.0008           & 258.7260$\pm$0.0004   \\
             Mean anomaly, $M$ (\degr)                         & = & 153.11119$\pm$0.00003         & 246.1$\pm$0.6       & 334.1$\pm$0.3       & 
                                                                     326.725$\pm$0.014             &  92.04$\pm$0.06       \\
             Perihelion, $q$ (AU)                              & = &   0.77453675$\pm$0.00000005   &   0.7802$\pm$0.0002 &   0.7781$\pm$0.0002 &   
                                                                       0.78201$\pm$0.00006         &   0.77555$\pm$0.00013 \\
             Aphelion, $Q$ (AU)                                & = &   1.88846206$\pm$0.00000002   &   1.9823$\pm$0.0014 &   1.9560$\pm$0.0012 &   
                                                                       1.9706$\pm$0.0004           &   1.9809$\pm$0.0009   \\
             MOID with the Earth (AU)                          & = &   0.0177805                   &   0.0151122         &   0.0447119         &   
                                                                       0.0000245866                &   0.00476759          \\
             Absolute magnitude, $H$ (mag)                     & = &  20.0                         &  26.4               &  24.2               &  
                                                                      30.5                         &  27.5                 \\ 
             data-arc span (d)                                 & = & 2115                          &  5                  &  16                 &   
                                                                     ---                           &  2                    \\ 
            \hline
         \end{tabular}
         \tablefoot{Values include the 1$\sigma$ uncertainty. The orbits of 454100, 2016~LR, 2018~BA$_{5}$ , and 2018~TU are 
                    referred to epoch JD 2458600.5, which corresponds to 0:00 on 2019 April 27 TDB; the orbit of 2018~LA is 
                    referred to epoch JD 2458271.5, which corresponds to 0:00 on 2018 June 2 TDB, and was produced by D. 
                    Farnocchia (J2000.0 ecliptic and equinox). Source: JPL's SBDB.
                   }
      \end{table*}
%
%---------------------------------------------------------------------------------------------------------------------------------
%

      In order to study the orbital evolution of 2018~LA and related objects, we have used a direct $N$-body code that implements a
      fourth-order version of the Hermite integration scheme (\citealt{1991ApJ...369..200M}; \citealt{2003gnbs.book.....A}). The 
      standard version of this software is publicly available from the Institute of Astronomy of the University of Cambridge 
      web site.\footnote{\href{http://www.ast.cam.ac.uk/~sverre/web/pages/nbody.htm}
      {http://www.ast.cam.ac.uk/$\sim$sverre/web/pages/nbody.htm.}} Practical details of this code, of solar system integrations, 
      and of our overall approach and physical model are described in \citet{2012MNRAS.427..728D} and \citet{2018MNRAS.473.3434D}. 
      Uncertainties are included in our analysis by applying the covariance matrix methodology described in \citet{2015MNRAS.453.1288D}. 
      The covariance matrices necessary to generate initial conditions have been obtained from JPL's SBDB, which is also the 
      source of other input data and ephemerides. Averages, standard deviations, interquartile ranges (IQRs), and other statistics 
      have been calculated in the usual way (see e.g. \citealt{2012psa..book.....W}).

   \section{Matching paths}
\begin{figure}
        \centering
        \includegraphics[width=\linewidth]{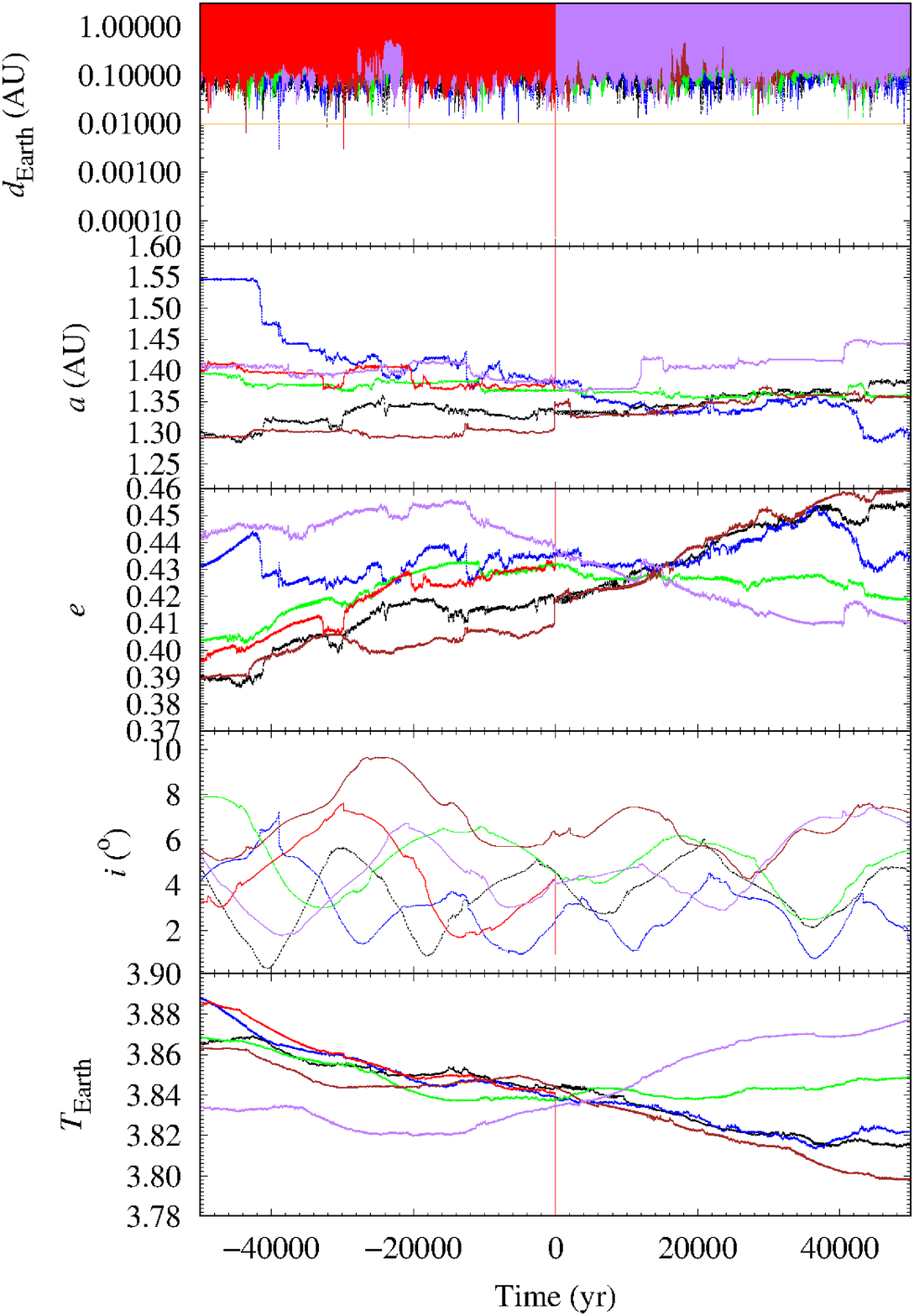}
        \caption{Evolution of the distance from the Earth to relevant NEOs (top panel) ---the value of the Hill radius of the 
                 Earth, 0.0098~AU, is plotted in orange--- semi-major axis (second to top panel), eccentricity (middle papel), 
                 inclination (second to bottom panel), and Tisserand parameter with respect to the Earth (bottom panel) of the 
                 nominal orbits of (454100) 2013~BO$_{73}$ (black), 2016~LR (blue), 2018~BA$_{5}$ (green), 2018~GU$_{11}$ (brown), 
                 2018~TU (purple), and 2018~LA (red). The zero-point in time corresponds to epoch JD~2458200.5~TDB, 2018 March 23.
                }
        \label{evolution}
     \end{figure}

\begin{figure}
        \centering
        \includegraphics[width=\linewidth]{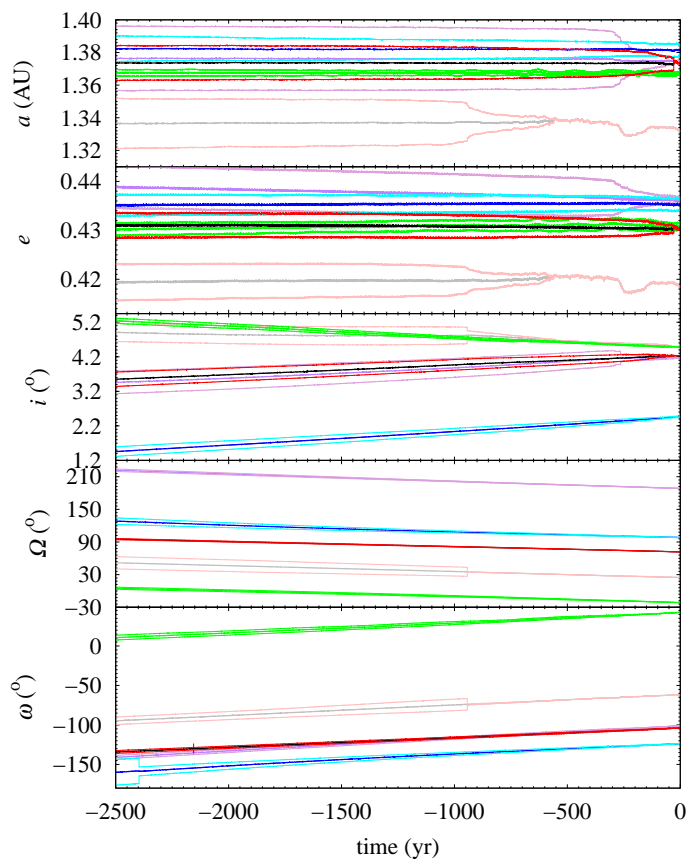}
        \caption{Evolution backwards in time of the dispersions of the values of the orbital elements of (454100) 2013~BO$_{73}$ 
                 (grey/pink), 2016~LR (blue/cyan), 2018~BA$_{5}$ (green/lime), 2018~TU (purple/lilac), and 2018~LA (black/red): 
                 semi-major axis (top panel), eccentricity (second to top panel), inclination (middle panel), longitude of the 
                 ascending node (second to bottom panel), and argument of perihelion (bottom panel). Average values are displayed 
                 as thick curves and their ranges (1$\sigma$ uncertainties) as thin curves. This figure shows results for 500 
                 control orbits of each object; initial positions and velocities have been computed using the covariance matrix. 
                 The zero-point in time is as in Fig.~\ref{evolution}.
                }
        \label{disper}
     \end{figure}

      Figure \ref{evolution} shows the orbital evolution of the nominal orbit (second column from the right in Table \ref{elements}) 
      of 2018~LA (in red) prior to becoming a fireball. Although our current study is based on the latest public data, the 
      dynamical behaviour of 2018~LA is quite sensitive to initial conditions and its past orbital evolution diverges quickly for 
      the various solutions released by JPL's SBDB (compare with Fig. 1 in \citealt{2018RNAAS...2b..57D,2018RNAAS...2c.131D}). 
      However, when the uncertainties in the orbit determination are included in the calculations via the covariance matrix, the 
      values of the orbital elements of 2018~LA remain well confined within a relatively narrow region of the NEO orbital 
      parameter space (see dispersion ranges in Fig.~\ref{disper} as red curves and average values as black ones).  
%
%---------------------------------------------------------------------------------------------------------------------------------
%
%
%---------------------------------------------------------------------------------------------------------------------------------
%

      As pointed out above, meteoroids like 2008~TC$_{3}$, 2014~AA, or 2018~LA might have been fragments of larger members of the 
      NEO population and none of the various processes that can lead to the production of such fragments are likely to spawn just 
      one meteoroid per event. Therefore, when 2018~LA was released, other debris may have followed. If this happened during the 
      relatively recent past, it may still be possible to identify putative siblings of 2018~LA and even plausible candidates from 
      which the observed meteoroids may have originated (see e.g. \citealt{2018ApJ...864L...9Y}). Following the approach discussed 
      in detail by \citet{2018MNRAS.473.3434D}, we have used the $D$-criteria -- which are metrics used to decide whether two 
      orbits bear some resemblance --  to search the known NEOs for objects that could be dynamically similar to 2018~LA. First, we 
      have selected NEOs with $D_{\rm LS}$ and $D_{\rm R} < 0.05$ (the lower the relative $D$ value, the more similar the 
      orbits are) and then studied their dynamical evolution using $N$-body integrations; $D_{\rm LS}$ in the form of Eq.~(1) 
      in \citet{1994ASPC...63...62L} and the $D_{\rm R}$ in the form of Eq.~(30) in \citet{1999MNRAS.304..743V}. We must 
      emphasize that having similar orbits in terms of the values of the osculating orbital elements is not enough to claim a 
      dynamical connection, the orbital evolutions over a suitable time interval must be consistent as well. 

      We have found well over two dozen objects whose current orbits resemble that of 2018~LA in terms of size ($a$), shape ($e$), 
      and orientation in space ($i$, $\Omega$ and $\omega$), but the final sample does not depend on $\Omega$ and $\omega$ due to 
      the mathematical forms of $D_{\rm LS}$ and $D_{\rm R}$. Out of this sample, some of them follow orbital evolutions that 
      compare well with that of 2018~LA. In Fig.~\ref{evolution}, we highlight the cases of (454100) 2013~BO$_{73}$, 2016~LR, 
      2018~BA$_{5}$, 2018~GU$_{11}$ , and 2018~TU, which are amongst the objects with the least uncertain orbit determinations. Although the 
      current orbit of 2016~LR (see nominal orbit in Table \ref{elements}) has the lowest values of the $D$-criteria, its past 
      evolution (in blue in Fig.~\ref{evolution}) is less similar to that of 2018~LA than those of 454100 (in black) or 
      2018~BA$_{5}$ (in green, see nominal orbit in Table \ref{elements}). When considering the uncertainties in the orbit 
      determinations (see Table \ref{elements}), 2018~TU appears to be the closest dynamical relative of 2018~LA, if the evolution 
      in $\Omega$ is not taken into account (see Fig.~\ref{disper}, with 2018~LA in black/red and 2018~TU in purple/lilac).

      The case of the potentially hazardous asteroid (PHA) 454100 deserves special consideration (see nominal orbit in Table 
      \ref{elements}) because it has a size of 550~m ($H=20$~mag, see Table 4 in \citealt{2016AJ....152...63N}) and a minimum 
      orbit intersection distance (MOID) to the Earth of 0.0179~AU. Asteroids 2016~LR and 2018~BA$_{5}$ are smaller than 
      454100, but larger than 2018~LA at $H=26.4$~mag and 24.2~mag, respectively. Other small NEOs with orbits similar to that of 
      2018~LA are 2018~GU$_{11}$ (in brown in Fig.~\ref{evolution}, $a=1.350\pm0.005$~AU, $e=0.420\pm0.003$, $i=6\fdg35\pm0\fdg04$, 
      $\Omega=203\fdg0184\pm0\fdg0011$, $\omega=77\fdg06\pm0\fdg06$, $H=28.4$~mag, and a MOID with the Earth of 0.0039~AU), 
      2018~TU (in purple in Fig.~\ref{evolution}, see nominal orbit in Table \ref{elements}), and 2018~UA ($a=1.3898\pm0.0009$~AU, 
      $e=0.4473\pm0.0003$, $i=2\fdg642\pm0\fdg005$, $\Omega=205\fdg6842\pm0\fdg0003$, $\omega=255\fdg21\pm0\fdg03$, $H=30.1$~mag, 
      and a MOID with the Earth of 0.000186~AU), all of them recent discoveries with rather uncertain orbit determinations 
      (particularly 2018~UA, with a data-arc span of 6.78 h).
%
%---------------------------------------------------------------------------------------------------------------------------------
%
%
%---------------------------------------------------------------------------------------------------------------------------------
%

      We have repeated the analysis shown in Fig.~\ref{disper} for 2018~LA (black/red curves) in the case of several of the 
      relevant minor bodies mentioned before and found that 454100 represents a promising match in the search for a candidate to 
      be the parent body of 2018~LA as it is the largest of the sample. The grey and pink curves in Fig.~\ref{disper} show the 
      results of the simulations, and the orbits of 454100 and 2018~LA are fully consistent in terms of size and shape when 
      3$\sigma$ deviations are taken into account. The orientation in space is somewhat different, but this opens the possibility 
      of allowing truly close fly-bys between these two objects. We have performed 10$^5$ $N$-body experiments integrating the 
      orbits of 454100 and 2018~LA backwards in time for 2000~yr and found that encounters at distances below 5000~km are possible; 
      in fact, nearly 11.5\% of the experiments produce an encounter at a mutual separation below one Lunar distance (384\,402~km). 
      These encounters take place at perihelion, at about 0.79~AU (median value) from the Sun and in the neighbourhood of 
      Venus, during the time window 500--1500~yr ago, and at relative velocities close to 2.6~km~s$^{-1}$.
%
%---------------------------------------------------------------------------------------------------------------------------------
%
      \begin{figure}
        \centering
         \includegraphics[width=\linewidth]{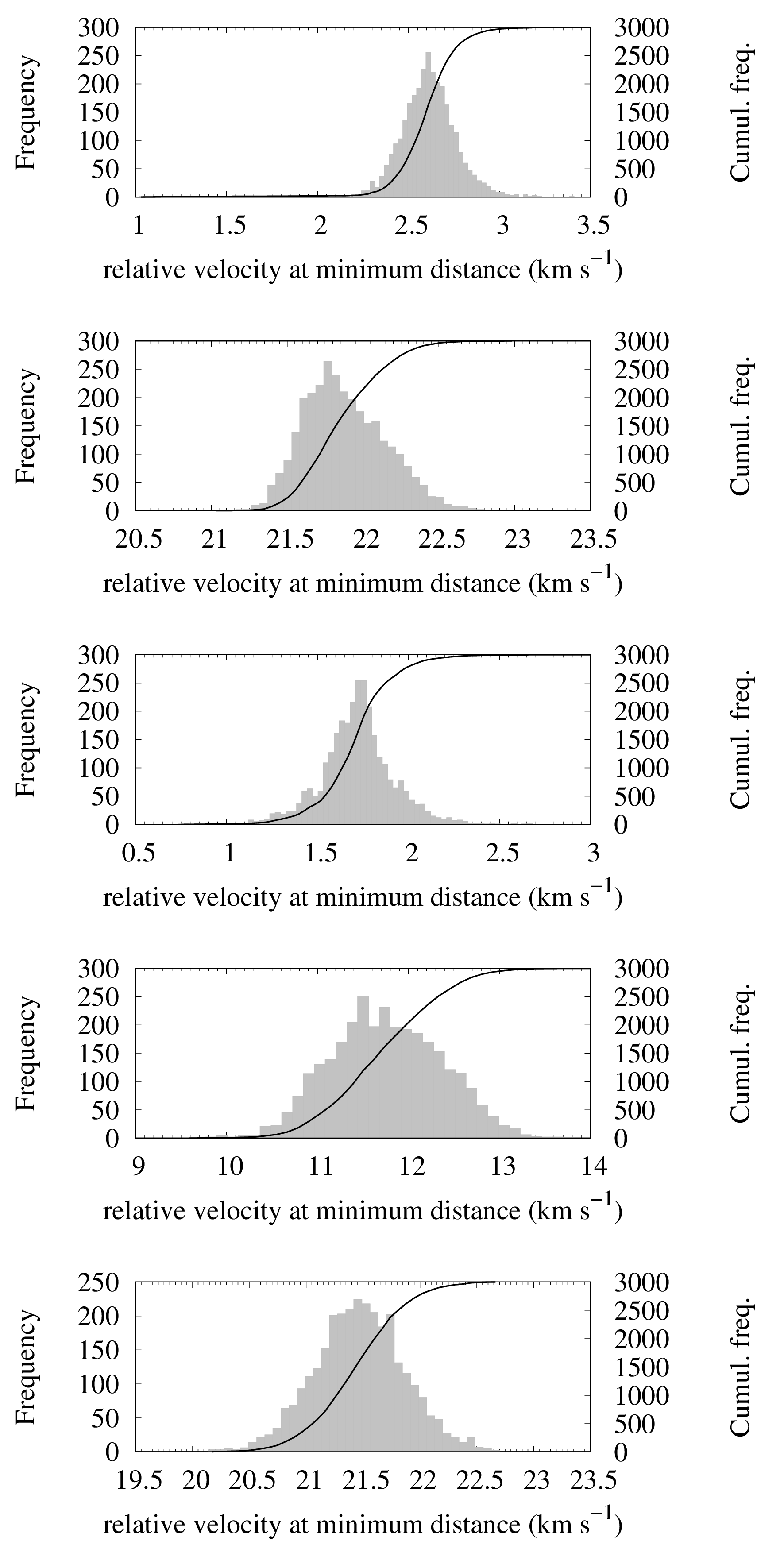}
         \caption{Velocity distribution between two virtual NEOs following control orbits statistically compatible with those of
                  (454100)~2013~BO$_{73}$ and 2018~LA (top panel) ---in the histogram, the bin width is 0.025~km~s$^{-1}$, with 
                  IQR = 0.194~km~s$^{-1}$ and a mean value of 2.6$\pm$0.2~km~s$^{-1}$; 454100 and 2018~TU (second to top panel) 
                  ---bin width is 0.053~km~s$^{-1}$, with IQR = 0.381~km~s$^{-1}$ and a mean value of 21.9$\pm$0.3~km~s$^{-1}$; 
                  2016~LR and 2018~LA (middle panel) ---bin width is 0.029~km~s$^{-1}$, with IQR = 0.212~km~s$^{-1}$ and a mean 
                  value of 1.8$\pm$1.9~km~s$^{-1}$; 2018~BA$_{5}$ and 2018~LA (second to bottom panel) ---bin width is 
                  0.119~km~s$^{-1}$, with IQR = 0.859~km~s$^{-1}$ and a mean value of 11.8$\pm$1.3~km~s$^{-1}$; 2018~LA and 
                  2018~TU (bottom panel) ---bin width is 0.071~km~s$^{-1}$, with IQR = 0.513~km~s$^{-1}$ and a mean value of 
                  21.5$\pm$0.4~km~s$^{-1}$.
                 }
         \label{YB}
      \end{figure}
%
%---------------------------------------------------------------------------------------------------------------------------------
%

      The relative velocity at minimum separation between two virtual NEOs following control orbits statistically compatible with 
      those of 454100 and 2018~LA follows the distribution shown in Fig.~\ref{YB}, top panel. These values are comparatively low 
      when considering collisions between members of the general NEO population (see e.g. Fig.~16 in \citealt{2018MNRAS.473.3434D}). 
      All these properties might be compatible with a scenario in which 2018~LA may have been released after a relatively recent 
      impact on 454100 but, how often might such an episode occur? Are there any recent observational data compatible with 
      present-day shattering impacts? The answer is that such collisions are unusual, but definitely not uncommon; asteroid 
      (596)~Scheila 1906~UA experienced a sub-critical impact in 2010 December, the impact velocity was probably close to 
      5~km~s$^{-1}$ (see e.g. \citealt{2011ApJ...733L...3B}; \citealt{2011ApJ...740L..11I}; \citealt{2011ApJ...733L...4J}), 
      which is the typical value across the main asteroid belt.

      However, the relative orbital orientation has an effect on the velocity distribution of collisions between objects following 
      2018~LA-like orbits. Figure~\ref{YB} shows the results of numerical experiments -- similar to those of 454100 and 2018~LA -- for 
      relevant pairs of NEOs following 2018~LA-like orbits. Gentle approaches are possible as in the case of 2016~LR and 2018~LA 
      (Fig.~\ref{YB}, middle panel), but high-speed fly-bys are also observed as in the case of 454100 and 2018~TU (Fig.~\ref{YB}, 
      second to top panel). Encounters between 2016~LR and 2018~LA take place at perihelion and within the time window discussed 
      for 454100 and 2018~LA. For 454100 and 2018~TU, nearly 5\% of the experiments produce an encounter at a mutual separation 
      below one Lunar distance, which resembles the behaviour observed in the case of 454100 and 2018~LA, but at much higher 
      relative speed, 21.9$\pm$0.3~km~s$^{-1}$. As for the dispersions in the orbital evolutions of these NEOs, Fig.~\ref{disper} 
      shows additional results that confirm that similar orbits and orbital evolutions, see the cases of 2018~LA and 2018~TU, can 
      still produce high-speed encounters. At such high speeds, 21.5$\pm$0.4~km~s$^{-1}$, catastrophic fragmentations are indeed 
      possible. 

   \section{Related meteor activity?}
      If 2018~LA was a fragment -- or a fragment of a fragment -- of (454100) 2013~BO$_{73}$, it is unlikely to be alone: a 
      sizeable stream of debris may occupy the same region of the orbital parameter space. If this hypothesis is correct, some 
      evidence in the form of meteors coming from approximately the same sky position around the same time of the year should 
      exist. The ephemerides show that the Botswana fireball of 2018 June 2 came approximately from $(\alpha_o, 
      \delta_o)$$\sim$$(243.6\degr, -10.5\degr)$, which is in the constellation of Scorpius, and had a velocity of 
      $\sim$17~km~s$^{-1}$. We have searched the available literature on meteor activity (e.g. \citealt{2006mspc.book.....J}) and 
      found one meteor shower that might be related to 2018~LA: the $\chi$-Scorpiids (\citealt{1973Icar...18..253S}). Activity for 
      this meteor shower is observed from May 28 to June 5 with a periodicity of 2.7 yr and \citet{1973Icar...18..253S} gives 
      a radiant of $(243.4\degr\pm1.9\degr, -10.8\degr\pm1.0\degr)$, a geocentric velocity of 19.9~km~s$^{-1}$, and mean orbital 
      elements referred to the equinox 1950.0 ---$a=1.943$~AU, $q=0.639\pm0.013$~AU, $e=0.671\pm0.017$, $i=6.8\degr\pm0.7\degr$, 
      $\Omega=65.9\degr\pm2.7\degr$, and $\omega=265.4\degr\pm1.6\degr$. \citet{1973Icar...18..253S} includes the 
      $\chi$-Scorpiids as a member of the Great Cluster of meteor showers, which he considers as the greatest concentration of 
      possibly related meteoroid streams. \citet{2013A&A...556A..25B} gives a list of NEOs that may be associated with the 
      $\chi$-Scorpiids and parameters for the meteor shower similar to those in \citet{1973Icar...18..253S}. 

      From the values in Table \ref{elements} and the observed radiant and velocity of the Botswana fireball, we can conclude that 
      the $\chi$-Scorpiids are a very good match in terms of timing (within the year), radiant and associated velocity, and 
      orientation in space, but the shape of the theoretical orbit is significantly more elongated than those of 454100, 2018~LA, 
      and related objects. Although the $\chi$-Scorpiids seem to be a promising match, the fact that meteor activity has long been 
      recorded from that area of the sky at the right time of the year lends support to the hypothesis that 2018~LA could be part 
      of a stream of debris with a possible origin in perhaps 454100. The recent close fly-by by 
      2018~UA\footnote{\href{https://www.minorplanetcenter.net/mpec/K18/K18U19.html}{https://www.minorplanetcenter.net/mpec/K18/K18U19.html.}} 
      at just 0.04 Lunar distances on 2018 October 19 (fourth-closest pass by the Earth on record, other than impacts), a 
      $H=30.1$~mag meteoroid that now follows an orbit very similar to that of 454100 (i.e. it matches the criterion discussed 
      above) suggests that the presence of dynamically coherent debris cannot be discarded in this case. Having discovered one 
      impactor and one very close passer-by with similar osculating orbital elements within about four months of each other may 
      be more than just a coincidence. Interestingly, the pre-encounter orbit of 2018~UA is given by JPL's SBDB as $a=1.930$~AU, 
      $e=0.550$, $i=6.4\degr$, which is close to that of the $\chi$-Scorpiids in \citet{1973Icar...18..253S}. 

   \section{NEO model predictions}
      If 2018~LA is associated with periodic meteor activity, this must be accompanied by a larger-than-expected fraction of 
      observed minor bodies following trajectories resembling those of 2018~LA or (454100) 2013~BO$_{73}$. If the hypothetical 
      ensemble of small bodies, which  are the source of the observed meteor activity, are the result of a relatively recent 
      fragmentation event of any kind, such a feature cannot be present in synthetic, debiased data from a NEO population model. 
      By comparing observational and synthetic data we may be able to understand better the circumstances surrounding 2018~LA's 
      impact. 

      The NEO orbit model developed within the framework of the Near-Earth Object Population Observation Program (NEOPOP) and 
      described by \citet{2018Icar..312..181G} is the state-of-the-art tool that can make such a comparison possible; this 
      new four-dimensional model of the NEO population describes debiased steady-state distributions of $a$, $e$, $i$, and $H$ in 
      the range $17 < H < 25$. When generating synthetic orbits from the NEO model, the two additional angular elements, $\Omega$ 
      and $\omega$, are drawn from flat distributions. It is, however, well known that the unbiased distributions in $\Omega$ and 
      $\omega$ are not flat and also that selection effects significantly affect the distributions of these angular elements (see 
      e.g. \citealt{2014Icar..229..236J}). \citet{2018Icar..312..181G} have estimated that the error arising from the 
      assumption of flat distributions is around 5\% or less for the debiased NEO population. The software for generating a 
      realization of the model is publicly available\footnote{\href{http://neo.ssa.esa.int/neo-population}{http://neo.ssa.esa.int/neo-population.}} 
      and the NEO model has already been used to complete several recent NEO studies that reproduce observed distributions of NEO 
      properties (\citealt{2016Natur.530..303G}, \citeyear{2017A&A...598A..52G}; \citealt{2018Icar..311..271G}). In addition, its 
      results are consistent with an independent analysis carried out by \citet{2017Icar..284..416T}.

      We have used the list of NEOs currently catalogued (as of 2018 November 11, 14\,439 objects with $H<25$~mag) and found nine 
      objects following 454100-like orbits and 17 pursuing 2018~LA-like ones (applying the $D$ criteria discussed above). The 
      orbit model, for a sample of the same size (i.e. 14\,439 objects), predicts 4.6$\pm$1.9 and 5.1$\pm$2.2, respectively, 
      which represent excess deviations of 2.3$\sigma$ and 5.4$\sigma$, respectively, of the numbers actually observed with 
      respect to those predicted: averages and standard deviations from 50 subsamples (of 14\,439 synthetic objects) of the 
      orbit model. This suggests that an excess of small bodies following trajectories resembling those of 2018~LA or (454100) 
      2013~BO$_{73}$ may exist and, if true, this is consistent with our hypothetical scenario in which 2018~LA and 454100 could 
      be related and associated with periodic meteor activity. It is, however, possible that current observational data may be 
      biased in favour of this type of orbit simply because they tend to pass closer to our planet (see the case of 2018~UA 
      above) and this fact makes them easier to discover. 

      The NEOPOP software predicts the existence of over 800\,000 NEOs with $H<25$~mag, but the size of the observed sample of 
      NEOs in the same range of absolute magnitudes is 14\,439. Therefore, we can put the completeness level of the overall NEO 
      sample in this magnitude range at $\sim$1.8\%. Following the same reasoning and for 454100-like orbits, the NEOPOP software 
      predicts 237 NEOs but the number actually observed is nine; therefore, the completeness level in this case could be $\sim$3.8\%. 
      For 2018~LA-like orbits, the NEOPOP software predicts 251 NEOs for an actual observational tally of 17, which gives a 
      completeness level of $\sim$6.8\%. In both cases, the completeness level is higher than the total one so we can assume that 
      NEOs following 454100-like or 2018~LA-like orbits are indeed easier to observe than the average NEO. However, objects in 
      orbits similar to that of 2018~LA are nearly twice as likely to be observed than those following 454100-like paths. These 
      results have been obtained assuming that $D_{\rm LS}$ and $D_{\rm R} < 0.05$, as we did in previous sections, and show that 
      extracting a subsample of 14\,439 synthetic NEOs from the model does not provide an orbit distribution that can be 
      meaningfully compared with the observed distribution of NEOs, because selection effects have not been accounted for. However, 
      subsamples generated using the NEOPOP sofware are expected to be free from observational biases for the most part, though not 
      fully for $\Omega$ and $\omega$ (see above).

      Let us assume that the null hypothesis is that the sample of NEOs linked to a given one ($D_{\rm LS}$ and $D_{\rm R}<0.05$) 
      can be detected from the NEO orbit model; in other words, the null hypothesis is that a similar cluster of orbits as 
      found around 2018~LA and 454100 would be found amongst synthetic NEOs. Although the empirical sample is affected by various 
      observational biases, we want to study how strongly the data can reject (or not) the null hypothesis. Following the ideas 
      discussed by \citet{1970smrw.book.....F}, here we apply a randomization test for a monovariate two-sample problem. This 
      non-parametric statistical test will tell us if the excesses found above are statistically significant. Our test statistics 
      is the difference between the observed number of NEOs with orbits similar to the one given and the predicted number using 
      synthetic data. As pointed out above, the empirical sample includes 14\,439 NEOs with $H<25$~mag and using the previously 
      mentioned criterion for similarity we obtain that, empirically, nine NEOs follow 454100-like orbits and 17 follow 2018~LA-like 
      orbits; the predictions using synthetic data are 4.6$\pm$1.9 and 5.1$\pm$2.2, respectively. Therefore, the respective 
      differences (our test statistics) are 4.4$\pm$1.9 and 11.9$\pm$2.2. In order to implement the randomization test, we  
      used one instance of the NEO orbit model (over 800\,000 NEOs) and extracted two random samples, each one including 14\,439 
      synthetic NEOs; then found for both samples how many synthetic NEOs had 454100-like orbits and how many had 2018~LA-like 
      orbits, and computed the differences. We repeated this experiment 10\,000 times to obtain the distribution of differences. 
%
%---------------------------------------------------------------------------------------------------------------------------------
%
      \begin{figure}
        \centering
         \includegraphics[width=\linewidth]{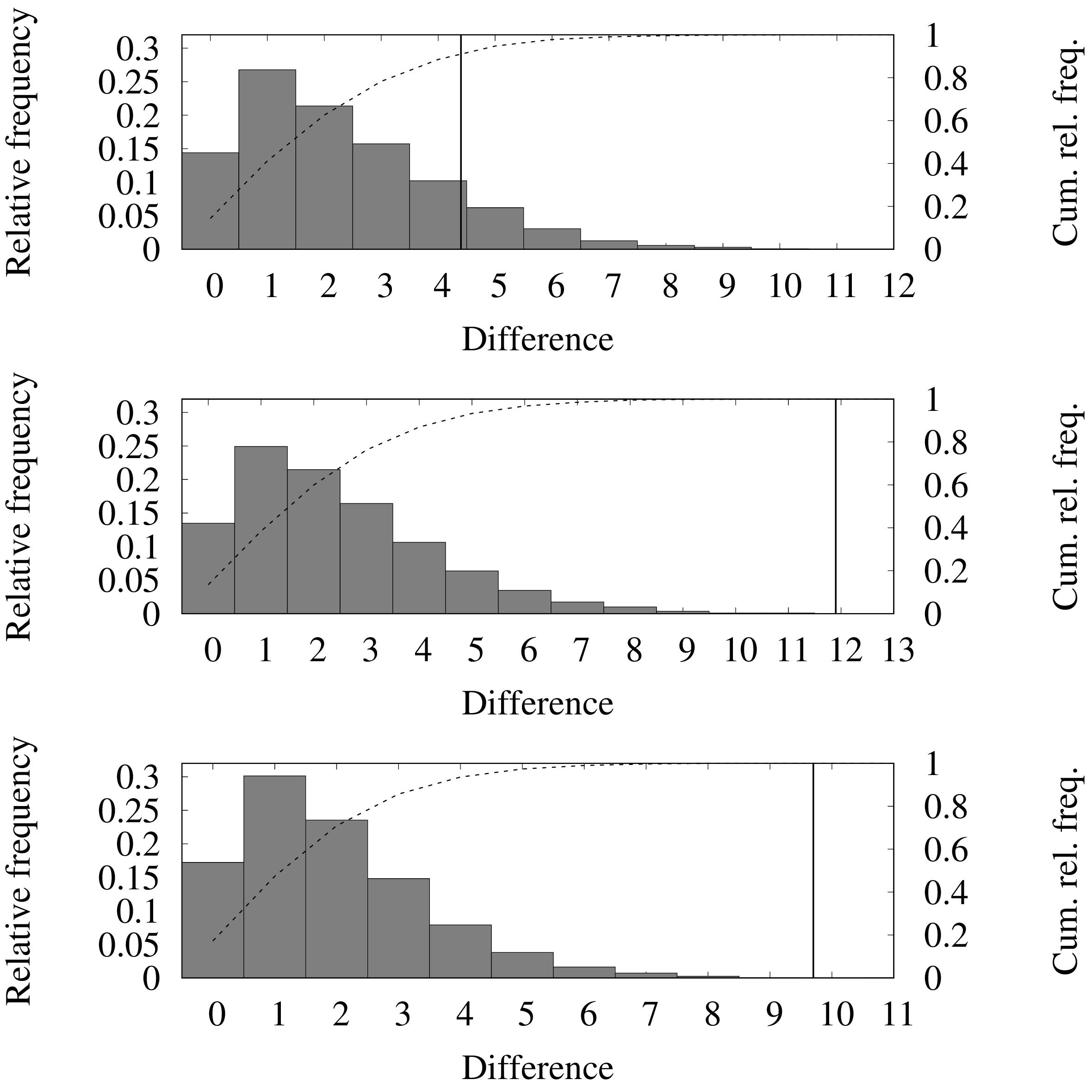}
         \caption{Results of the randomization test described in the text. The observed average differences betwen empirical and 
                  synthetic counts are represented by vertical lines. The top panel shows results for (454100) 2013~BO$_{73}$-like 
                  orbits; the middle one shows those of 2018~LA-like orbits; the bottom panel shows those of 2018~UA-like orbits. 
                 }
         \label{random}
      \end{figure}
%
%---------------------------------------------------------------------------------------------------------------------------------
%

      The results of these experiments are plotted in the form of histograms in Fig.~\ref{random} together with the observed 
      average values of the test statistics (as vertical black lines). From the figure, the excess of 454100-like orbits (top 
      panel) may not be out of the ordinary, but the one observed for 2018~LA-like orbits (middle panel) is significant as it is well 
      separated from the bulk of the distribution. Our results show that the probability of getting a difference $>2.5$ for 
      454100-like orbits is 0.3746 and that of getting a value $>4.4$ is 0.1150 (see Fig.~\ref{random}, top panel). Therefore and 
      for 454100-like orbits, the null hypothesis cannot be rejected at any meaningful level of significance. In sharp contrast 
      and for 2018~LA-like orbits, the probability of getting a difference $>9.7$ is 0.0017 and that of obtaining $>11.9$ is 
      0.0003 (see Fig.~\ref{random}, middle panel), which means that the null hypothesis can be safely rejected for these orbits. 
      In summary, the answer to the original question  -- how likely is such a large number of NEOs moving in 2018~LA-like orbits if 
      those orbits come from a distribution compatible with the one provided by the NEO orbit model? -- is a robust ``very 
      unlikely". As the 2018~LA-like orbits have assumedly evolved out of 454100-like ones within the context of our hypothesis, 
      relatively stable orbits turn into those that may lead to an impact, passing regularly closer to our observation point on 
      the Earth. If we focus on 2018~UA-like orbits (the close passer-by with very uncertain orbit) instead, we have an empirical 
      tally of 13, a synthetic one of 3.3$\pm$1.8, and a test statistics of 9.7$\pm$1.8; a similar set of experiments shows that 
      the probability of getting a difference $>7.9$ is 0.0027 and that of obtaining $>9.7$ is 0.0001 (see Fig.~\ref{random}, 
      bottom panel), which indicate that the excess of 2018~UA-like orbits is also statistically significant.

   \section{Discussion}
      The study of possible dynamical and/or genetic relationships amongst members of the NEO population has implications in 
      knowledge areas as diverse as the evolution of the NEO population, the study of the near-Earth space environment, in situ 
      exploration and commercial applications of NEOs, and planetary defense (see e.g. \citealt{2012Natur.485..549E}; 
      \citealt{2014ApJ...785L...4H}; \citealt{2017JGRE..122..789M}). Our extensive calculations indicate that within the NEO
      orbital parameter space both low-speed (relative velocity under 2~km~s$^{-1}$) and high-speed (above 20~km~s$^{-1}$) 
      collisions are possible; that is, cratering and catastrophic impacts are both feasible amongst members of dynamically similar NEO 
      groups. This could be considered as an additional piece of evidence in favour of fragmentation taking place in near-Earth 
      space; a more robust example has been discussed by \citet{2019MNRAS.483L..37D}, but in that case fission via the YORP 
      mechanism (see e.g. \citealt{2011Icar..214..161J}), not actual impacts, is thought to be the cause. 
      
      Our statistical tests confirm that the presence of (454100) 2013~BO$_{73}$ is compatible with predictions from the NEO orbit 
      model used here, but that of 2018~LA (and 2018~UA) is probably not. However, and with the available data, we cannot tell 
      anything certain about the significance of the link between 454100 and 2018~LA, spectroscopic (i.e. compositional) 
      information is needed in order to do this properly. After analysing the relative dynamics of relevant NEOs following these 
      orbits, it cannot be discarded that most members of the group could be the result of collisional cascades that produce 
      multi-generation fragments as envisioned by, for example, \citet{1994IAUS..160..205F}. The dynamical landscape is consistent 
      with this scenario; fragments of fragments can easily be produced under those conditions. 

      As pointed out above, the orbit model used here has intrinsic limitations but $a$, $e,$ and $i$ define the Tisserand 
      parameter (see e.g. \citealt{1999ssd..book.....M}) that is approximately conserved during the close encounters of these 
      objects with Venus and the Earth--Moon system as shown in Fig.~\ref{evolution}, bottom panel. Even if the true distributions 
      of $\Omega$ and $\omega$ have been disregarded in our analysis, the essence of the dynamical processes driving the orbital
      evolution of the relevant NEOs has been captured by our extensive simulations and statistical analyses, and their overall
      conclusions can be considered as sufficiently robust. It is also worth mentioning that the orbits of 2018~LA and related 
      objects are very typical of Earth's impactors as seen in Fig. 1 of \citet{2009Icar..203..472V}, which strongly suggests
      that they must be preferentially removed (relative to other NEO orbits), yet they seem to be overabundant. Assuming that
      this is not the result of some hidden and strong observational bias or selection effect, a natural scenario that explains 
      qualitatively what is observed is in the existence of collisional cascades still taking place within this dynamical group; 
      objects like 2018~LA are lost, but new impacts (within the group) replenish the losses (being in the form of impacts or 
      NEOs being scattered away from the dynamical neighbourhood of 2018~LA). 

   \section{Conclusions}
      In this paper, we have studied the pre-impact orbital evolution of 2018~LA, the parent body of the fireball observed over 
      South Africa and Botswana on 2018 June 2. The dynamical evolution of other, perhaps related, minor bodies has been explored 
      as well. This research has been carried out using the latest data, $N$-body simulations, a state-of-the-art NEO orbit model, 
      and statistical analyses. Our conclusions can be summarized as follows.
      \begin{enumerate}[(i)]
         \item We have identified a group of known NEOs that follow orbits similar to that of 2018~LA and whose dynamical 
               behaviour bears some resemblance to that of 2018~LA, prior to its impact. The largest member of this group is 
               (454100) 2013~BO$_{73}$.
         \item Extensive calculations show that the pair 454100--2018~LA may have experienced fly-bys at very close range in the 
               past. Based on this fact, we speculate that 2018~LA might have its origin in a relatively recent shattering event 
               that perhaps affected 454100. This putative fragmentation episode may have produced additional debris that might 
               still be identifiable in the available data.
         \item Periodic meteor activity, $\chi$-Scorpiids, that might be consistent with the properties of the radiant of 2018~LA 
               has been found recorded in the literature.
         \item We find a statistically significant excess of small bodies moving in orbits similar to that of 2018~LA when 
               comparing with predictions from a state-of-the-art NEO orbit model. This excess can be tentatively explained as a 
               result of collisional cascades operating within the dynamical group. 
      \end{enumerate}
      Future spectroscopic studies of 454100 (and other perhaps related NEOs) should be able to confirm if this PHA could be the 
      source of 2018~LA and other NEOs (by finding, or not, a chemical composition consistent with that of the recovered 
      meteorites of 2018~LA), and perhaps other extraterrestrial materials that have been observed historically as coming from the 
      same general area of the sky in Scorpius.

   \begin{acknowledgements}
      The authors thank the referee, M. Granvik, for a constructive review that led to significant improvements in the 
      interpretation of our results within the context of the NEO orbit model used here, S.~J. Aarseth for providing one of the 
      codes used in this research, P. Mialle for comments on CTBTO infrasound recordings from the meteor blast, S. Deen for 
      comments on the overall organization of 2018~LA-like orbits, and  A.~I. G\'omez de Castro for providing access to computing 
      facilities. This work was partially supported by the Spanish `Ministerio de Econom\'{\i}a y Competitividad' (MINECO) under 
      grant ESP2015-68908-R. In preparation of this paper, we made use of the NASA Astrophysics Data System, the ASTRO-PH e-print 
      server, and the MPC data server.
   \end{acknowledgements}
 
   \bibliographystyle{aa}

\end{document}